\documentclass[structabstract]{aa}  
\usepackage{graphicx}
\usepackage{txfonts}
\usepackage{natbib}
\usepackage{aalongtable}
\bibpunct{(}{)}{;}{a}{}{,}
\newcommand{\nz}{\ensuremath{\langle N_z\rangle}}
\newcommand{\pv}{\ensuremath{P_V}}
\newcommand{\nv}{\ensuremath{N_V}}
\newcommand{\te}{\ensuremath{T_{\rm eff}}}
\newcommand{\bz}{\ensuremath{\langle B_z \rangle}}
\newcommand{\bs}{\ensuremath{\langle |B| \rangle}}
\newcommand{\kms}{km\,s\ensuremath{^{-1}}}

\begin{document}

\title{On the incidence of weak magnetic fields in DA white dwarfs\thanks{Based on observations collected at
    the European Organisation for Astronomical Research in the
    Southern Hemisphere, Chile, under observing programme 073.D-0516, 
    and obtained from the ESO/ST-ECF Science Archive Facility.}}

   \author{J. D. Landstreet  
     \inst{1,2} 
     \and S. Bagnulo
     \inst{1} 
     \and G. G. Valyavin 
     \inst{3} 
     \and L. Fossati 
     \inst{4} 
     \and S. Jordan
     \inst{5} 
     \and D. Monin 
     \inst{6} 
     \and G. A. Wade 
     \inst{7} }
   \institute{Armagh Observatory, College Hill, Armagh, BT61 9DG, 
     Northern Ireland, UK\\     
     \email{sba@arm.ac.uk}
     \and
     Department of Physics and Astronomy, The University of
     Western Ontario, London, Ontario, N6A 3K7, Canada\\
     \email{jlandstr@uwo.ca; jls@arm.ac.uk}
     \and
     Special Astrophysical Observatory, Nizhnij Arkhyz, Zelenchukskiy
     Region, Karachai-Cherkessian Republic, Russia 369167\\
     \email{gvalyavin@sao.ru}
     \and
     Argelander-Institut f\"{u}r Astronomie der Universit\"{a}t Bonn, 
     Auf dem H\"{u}gel 71, 53121 Bonn, Germany\\
     \email{lfossati@astro.uni-bonn.de}
     \and
     Astronomisches Rechen-Institut, Zentrum f\"ur Astronomie der 
     Universit\"at Heidelberg, M\"onchhofstr. 12-14, D-69120 
     Heidelberg, Germany\\
     \email{jordan@ari.uni-heidelberg.de}
     \and
     Dominion Astrophysical Observatory, Herzberg Institute of
     Astrophysics, National Research Council of Canada, 5071 West
     Saanich Road, Victoria, BC V9E 2E7, Canada\\
     \email{dmitry.monin@nrc.gc.ca}
     \and
     Department of Physics, Royal Military College of Canada,
     PO Box 17000, Stn Forces, Kingston, Ontario K7K 7B4, Canada\\
     \email{gregg.wade@rmc.ca}
     }
   \date{Received June 18, 2012; accepted July 29, 2012}

% \abstract{}{}{}{}{} 
% 5 {} token are mandatory
 
  \abstract
  {About 10\,\% of white dwarfs have magnetic fields with strength in 
   the range between about $10^5$ and $5\, 10^8$~G. It is not known
   whether the remaining white dwarfs are not magnetic, or if they
   have magnetic fields too weak to be detected with the techniques
   adopted in the large surveys. Information is particularly lacking
   for the cooler (and generally fainter) white dwarfs.}
  {We describe the results of the first survey specifically devised to
   clarify the detection frequency of kG-level magnetic fields in cool
   DA white dwarfs.}
 {Using the FORS1 instrument of the ESO VLT, we have obtained Balmer
   line circular spectropolarimetric measurements of a small sample of
   cool (DA6 -- DA8) white dwarfs.  Using FORS and UVES archive data,
   we have also revised numerous white dwarf field measurements
   previously published in the literature.}
 { We have discovered an apparently constant longitudinal magnetic
   field of $\sim 9.5$\,kG in the DA6 white dwarf \object{WD\,2105$-$820}.
   This star is the first weak-field white dwarf that has been
   observed sufficiently to roughly determine the characteristics of
   its field.  The available data are consistent with a simple dipolar
   morphology with magnetic axis nearly parallel to the rotation axis,
   and a polar strength of $\simeq 56$\,kG.  Our re-evaluation of the
   FORS archive data for white dwarfs indicates that longitudinal
   magnetic fields weaker than 10\,kG have previously been correctly
   identified in at least three white dwarfs. However, for one of
   these three weak-field stars (\object{WD\,2359$-$434}), UVES archive data
   show a $\sim 100$\,kG mean field modulus.  Either at the time of
   the FORS observations the star's magnetic field axis was nearly
   perpendicular to the line of sight, or the star's magnetic field
   has rather complex structure.  }
 {We find that the probability of detecting a field of kG strength in
   a DA white dwarf is of the order of 10\,\% for each of the cool and
   hot DA stars. If there is a lower cutoff to field strength in white
   dwarfs, or a field below which all white dwarfs are magnetic, the
   current precision of measurements is not yet sufficient to reveal
   it.}  
   \maketitle
%
%________________________________________________________________

\section{Introduction}

In 1970, a magnetic field was discovered in the peculiar white dwarf
(WD) Grw\,$+70\,8247$ = GJ\,472 \citep{Kempetal70}. The field strength
was eventually estimated to be of the order of 300\,MG
\citep{Greenstein84,WickramasingheFerrario88,Jordan92}. Since this
first detection of a magnetic field in a degenerate star, about 200
magnetic white dwarfs (MWDs) have been discovered
\citep{Kawkaetal07,Kulebietal09b}. It is found that about 10\,\% of
all single WDs have a magnetic field with a strength in the range
between hundreds of kG and hundreds of MG.

It is not at all clear how the magnetic fields in WDs originate, nor
what information they carry about the origin and evolution of
magnetism during stellar evolution. It is also not very clear yet how
these fields influence such phenomena as rotation periods or pulsation
of white dwarfs. Clearly, a broad observational base of data is
essential for understanding these issues.

The magnetic fields of WDs are sometimes variable with the stellar
rotation period, which when measurable is typically of the order of
hours or days \citep[e.g. ][]{Kawkaetal07}. It appears that MWDs may
often be somewhat more massive than the overall WD average mass of
about $0.6\,M_\odot$ \citep{Liebert88}, although fields are
occasionally found in relatively low-mass WDs. Most of the fields
known are in WDs of spectral type DA, a white dwarf classification
indicating that the optical spectrum shows only spectral lines of
hydrogen, and which generally identifies WDs with H-rich atmospheres.
This is at least partly a selection effect due to the fact that the
strong and sharp Balmer lines are particularly sensitive probes of
stellar magnetism, which in many cases can be easily detected in
low-dispersion spectra from surveys such as the Sloan Digital Sky
Survey \citep{Kulebietal09a,Kulebietal09b}.

The concentration of WD magnetic field strengths as a
function of $\log \bs$ \citep{Kulebietal09b} in the best-studied range
of 1--100\,MG has raised the question of whether there is a
cutoff field strength below which white dwarf fields do not occur
\citep[as is the case for Ap stars,][]{Auriereetal07}, or whether the
probability of detecting a field might rise sharply below a field
strength of some tens of kG.  Resolving this question confronts the
difficulty of detecting weak fields in such faint, broad-lined
objects, and our current knowledge of the low-field tail of the white
dwarf field strength distribution is limited mainly by instrumental
constraints. It is very difficult to obtain field measurements with
standard errors of less that about 10~kG without using the largest
available telescopes \citep[see e.g.][]{Valyavinetal06, Kawkaetal07}.
However, the study of available statistics by \citet{Liebertetal03}
and the survey by \citet{Aznaretal04} both suggest that the detection
rate for field weaker than a few tens of kG may be significantly
higher than the frequency of $\sim 10$\,\%, which characterises the
overall detection rate of stronger fields.

A further question of great interest is whether the magnetic fields of
WDs evolve with time, and if so, how they evolve. The searches for kG
fields reported so far
\citep{Aznaretal04,Valyavinetal06,Kawkaetal07,Jordanetal07} have
focussed almost entirely on the generally brighter hotter (and therefore
younger) white dwarfs. It is thus worthwhile to focus a survey on cooler
and older white dwarfs, and the higher detection probability predicted
by earlier work suggests that even a fairly small sample of such stars
may yield interesting results.

Thus, to increase the available information about the incidence of
weak fields, and to extend this information to include some older,
cooler white dwarfs, we have carried out a modest survey for fields in
DA WDs with effective temperatures \te\ below about
14\,000~K, aiming at obtaining field measurements with $\sim 1$\,kG
error bars. 

Recent work by \citet{Bagnuloetal12}, \citet{Jordanetal12}, and
\citet{Lanetal12} have shown that the results of some FORS1 surveys of
magnetic fields in various classes of stars were affected by spurious
detections, highlighting the need for a re-analysis of published data
for MWDs. Therefore, we have complemented the results of our own
survey with the revision of all FORS1 field measurements of WDs.

\section{New observations}\label{Sect_New_Observations}

White dwarfs with very strong fields can be identified via broad-band
circular polarimetry, as magnetic fields may produce circular
polarization of the continuum radiation at the level of 1 to a few\,\%
for fields of 10 MG or more. However, most MWDs have been detected by
observing the Zeeman effect in the Stokes $I$ and/or $V$ profiles of
spectral lines. 

For a 100~kG magnetic field, the $\pi - \sigma$ separation produced by
the Zeeman effect in optical spectral lines is $\sim 1$\,\AA. For DA
WDs, this is of the same order as the pressure broadening of the
Balmer line cores \citep{Koesteretal98}. This sets a lower limit to
the strength of the field that can be detected through intensity
measurements, since for a field strength $\la 30$\,kG, Zeeman
splitting no longer dominates over pressure broadening, and weak
splitting is difficult to distinguish from rotational line broadening.
Practically, most past surveys could firmly detect only fields with
$\bs \ga 50-100$\,kG, as at lower field strength the Zeeman splitting
in Stokes $I$ would be beneath the resolving-power limit of the
instrument, and/or swamped by noise.

In conclusion, for field with strength $\la 50$\,kG, the most
appropriate method for field detection is based on low-resolution,
high signal-to-noise ratio (S/N) measurements of the circular
polarization of spectral lines, which can be obtained with large
telescopes \citep{Landstreet92,Schmidt01}. Circular spectropolarimetry
of Balmer lines is the tool best suited for our field survey of DA
stars.

\begin{table*}
\caption{New longitudinal magnetic field measurements obtained with FORS1.}
\label{Tab:survey_obs}
\begin{center}
\begin{tabular}{llrlrcrrr@{$\,\pm\,$}lr}
  \hline\hline
  \multicolumn{2}{c}{Star names}&  $V$   & Spectral& \te &$\log g$&\multicolumn{1}{c}{MJD}   &$t_{\rm int}$&\multicolumn{2}{c}{\bz}& $\vert\bz/\sigma_{\rm \bz}\vert$ \\        
  \multicolumn{2}{c}{}       & (mag)  & type  & (K)   &      &        &    (s)    &\multicolumn{2}{c}{(G)}&                                \\
  \hline
  WD\,1425$-$811 & GJ 2108  & 13.0    & DA6:V & 12098 & 8.21 & 53137.044 &  960 &     488  &  1360 & 0.36 \\ %%% sp OK. meas c "all" has sig = 1056 G
  WD\,1733$-$544 & GJ 4012  & 15.8    & DA8:CV&  6165 & 7.23 & 53199.178 & 1664 &    4104  &  4390 & 0.94 \\ %%% Spectrum of this obj looks like A star, not
  WD\,1826$-$045 & G 21-16  & 14.5    & DA6   &  9057 & 7.99 & 53193.179 & 1920 & $-2705$  &  1530 & 1.76 \\ %%% sp OK. Meas c "all" has sigma = 1179 G
  WD\,1952$-$206 & LTT 7873 & 15.0    & DA6   & 13184 & 7.82 & 53251.088 & 2840 &     530  &  1180 & 0.45 \\ %%% sp OK. Meas c "all" has sigma = 960 G
  WD\,2105$-$820 & LTT 8381 & 13.5    & DA6   & 10794 & 8.19 & 53192.269 & 1760 &    9274  &  1375 & 6.75 \\ 
                 &          &         &       &       &     & 53193.278 & 1760 &   11423  &  995  & 11.47 \\ 
                 &          &         &       &       &     & 53197.294 &  880 &    8173  &  1630 & 5.02 \\ 
                 &          &         &       &       &     & 53199.317 &  880 &    9130  &  1490 & 6.12 \\ 
                 &          &         &       &       &     & 53227.209 & 1760 &    9770  &  845  & 11.59 \\ %%% sp OK. With "all" sigma = 755 G
  WD\,2115$-$560 & GJ 4191  & 14.3    & DA6   &  9625 & 8.01 & 53199.342 & 1664 & $-1367$  &  1065 & 1.28 \\ %%% sp OK. With "all" sigma = 895 G
                 &          &         &       &       &     & 53227.238 & 1664 &     132  &  955  & 0.14 \\ %%% sp OK. With "all" sigma = 772 G
  WD\,2151$-$015 & GJ 4236  & 14.5    & DA6   &\multicolumn{1}{c}{?}    
                          &\multicolumn{1}{c}{?}  & 53240.174 & 1840 &    3941  &  1910 & 2.08 \\ 
%%% I sp OK. V has big dip at 3850 A
                 &          &        &       &       &     & 53251.124 & 1840 & $-2027$  &  950  & 2.13 \\ %%% sp OK. H lines rather narrow. 
                 &          &        &       &       &     & 53252.120 & 1840 &  $-688$  &  1660 & 0.41 \\ 
%%% sp OK.
  WD\,2333$-$049 & G 157-82 & 15.9   & DA6   & 10608 & 8.04 & 53274.201 & 1704 &    5102  &  5550 & 0.92 \\ %%% sp OK but very noisy.
  \hline
\end{tabular}
\end{center}
\end{table*}

Our circular spectropolarimetric observations (programme ID
073.D-0516) were carried out in service mode during 2004 using FORS1
on the ESO VLT telescope Antu. Our survey draws randomly on a list of
nearby cool ($\te \la 14\,000$~K) WDs\footnote{Our programme was
  granted the status of ``filler'', and only a fraction of the
  observations originally planned were actually carried out.}. Targets
were observed using grism 600B with a 1.0-arcsec slit. Our FORS spectra
have a resolving power of about 830, and cover the wavelength window
from 3470 to 5880~\AA, thus including all the hydrogen Balmer
lines from H$\beta$ down to the series limit at about H9.

For each stellar observations, we typically obtained four integrations
with the quarter-wave plate rotated by $90^\circ$ between successive
exposures \citep[an observing procedure that makes it possible 
to eliminate a number of sources of measurement error to
first order -- see, e.g.,][]{Bagnuloetal09}.  Data reduction
and field measurements were performed as explained by
\citet{Bagnuloetal12}.  In particular, the mean line-of-sight magnetic
field \bz\ was obtained by using the relationship
\begin{equation}
V(\lambda) = -g_{\rm eff} C_{\rm Z} \lambda^2
\frac{{\rm d}I(\lambda)}{{\rm d}\lambda} \bz
\end{equation}
\citep{Landstreet82}, where $C_{\rm Z} = e/4\pi m c^2$, as a
correlation equation between the slope d$I$/d$\lambda$ of the
  local spectral intensity $I(\lambda)$, and the local circular
  polarisation $V(\lambda)$, pixel by pixel, as explained in detail
by \citet{Bagnuloetal12}.  However, in computing the slope of the
correlation between the value of $V/I$ with d$I/$d$\lambda$, sigma
clipping has now been introduced to remove outliers (mostly from
cosmic rays) that add noise but no real signal.  Since all of the
stars observed are DA stars, real magnetic signal is only found in the
H lines, and therefore field strengths were determined using only
these lines. The wavelength window to use for each line of each star
was set after visual inspection of the $I$ spectrum, to include all of
each line wing out to the point where the line slope decreases to
typical values produced by noise in the continuum.

For each observation we have also produced a null spectrum \nv, a
quantity computed by combining the circular polarisation spectra from
the four sub-exposures of the observation in such a way as to cancel
out the the real circular polarisation signal. The value of the null
spectrum is that it can reveal artefacts or systematic errors in the
data (due for example to cosmic rays). In a successful observation the
\nv\ spectrum should be featureless at the level of the photon noise,
and the magnetic field deduced from \nv\ should be consistent with
zero within its uncertainty. The computation and meaning of \nv\ are
discussed at length by \citet{Bagnuloetal09,Bagnuloetal12}.

The target list, observing log, and field measurments are given in
Table~\ref{Tab:survey_obs}, which provides: two names (cols. 1 and 2);
visual magnitude $V$ (col.\ 3); the spectral class (col.\ 4);
the effective temperature \te\ in K and the logarithm of the
gravity $g$ in cm\,s$^{-2}$, both taken from \citet{Lajoieetal07} or
\citet{Koesteretal09}; the modified Julian Date (MJD) of the midpoint
of each observation (col.\ 7); the total integration time $t_{\rm
  int}$ in sec (col.  8); the measured value of the mean longitudinal
field strength \bz\ and its standard error $\sigma_{\rm \bz}$ (col.\
9); and the significance of the detection, $\vert\bz/\sigma_{\rm
  \bz}\vert$. The survey comprises 15 individual measurements of eight
different stars, and required about 8\,h of telescope time in service
mode (out of 42\,h originally planned).

The observed WDs are quite faint (their magnitude ranges from $V\sim
13$ to $\sim 16$), and in some of them, the low values of \te\ lead
to rather weak Balmer lines. Nevertheless, it may be seen that the
precision sought for these measurements has been, to a considerable
extent, achieved: all but two of the 15 measurements have standard
errors in the range of 800 to 2000~G.

\section{Results}

\subsection{Detection of a kG field in WD\, 2105$-$820}\label{Sect_Detection}

Only one of the eight cool DA WDs of Table~\ref{Tab:survey_obs} shows
clear evidence of a magnetic field of kG strength, namely
WD\,2105$-$820 = GJ\,820.1 = LTT\,8381, which is a DA6 star with $\te
= 10800$\,K. This star had previously been flagged by \citet{Koesteretal98}
as potentially magnetic, on the basis of showing
excess broadening (and possibly Zeeman splitting) in the core of
H$\alpha$, although they point out that the observed broadening could
instead be due to rapid rotation with $v \sin i = 65$\,\kms.  For this
star, we have five \bz\ measurements with a typical (median) standard
error of about 1400\,G. Our five measurements reveal a longitudinal
field $\bz \approx +9500$\,G, with a $\sim 1200$\,G dispersion,
similar to the median measurement uncertainty. The significance of
the individual detections, $\vert\bz/\sigma_{\bz}\vert$, ranges from
about 5 to more than 10. Even with the problem of occasional outliers
among field measurements obtained with FORS1
\citep[see][]{Bagnuloetal12}, these detections are sufficiently
significant and numerous to allow us to conclude that the field is
certainly present.

\begin{figure}
\scalebox{0.44}{
\includegraphics*[0.9cm,5.6cm][21.5cm,25cm]{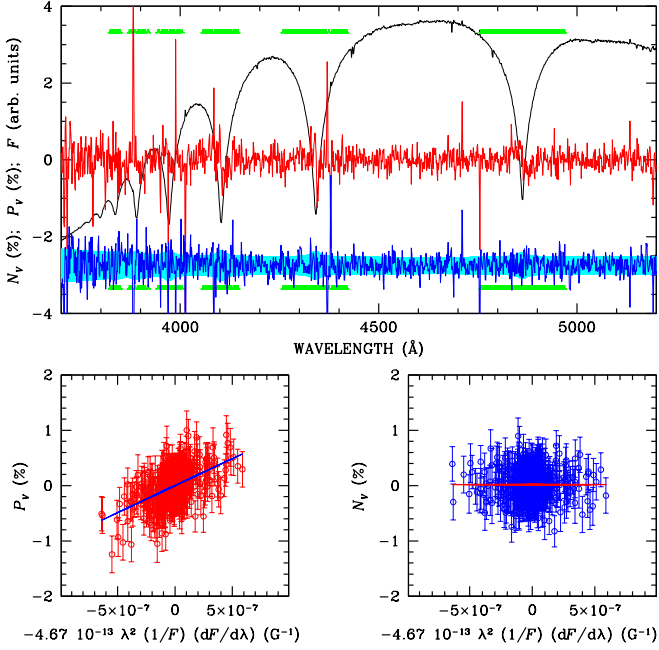}}\\
\caption{\label{Fig:wd2105-820-5} 
The observations of WD\,2105$-$820 obtained with FORS1 on 2004-08-10\,UT\,05:01 =
MJD 53227.209.  The
top panel shows the observed flux $F$ (black solid line, in arbitrary
units, and not corrected for the instrument response), the
$\pv = V/I$ profile (red solid line centred about 0), and the null profile
\nv\ (blue solid line, offset by $-2.75$\,\% for display purpose). The null
profile is expected to be centred about zero and scattered according
to a Gaussian with $\sigma$ given by the \pv\ error bars, which
are represented with light blue bars centred about
$-2.75$\,\%.  The regions used for field measurement are marked with green
bars above and below this spectrum. The slope of the interpolating lines in the bottom panels
provides the mean longitudinal field from \pv\ (left bottom panel) and
from the null profile (right bottom panel) both calculated using only the H
Balmer lines. The corresponding \bz\ and \nz\ values are $9770
\pm 843$\,G and $-11 \pm 868$\,G, respectively.
}
\end{figure}

The $I$, $V/I$ and $N_V$ spectra of one observation of WD\,2105$-$820
are shown in Fig.~\ref{Fig:wd2105-820-5}.  One can clearly see the
weak S-shaped excursions around zero in $V/I$ at the positions of
several of the Balmer line cores that reveal the presence of the field
of this star.  Because our observations have quite low spectral
resolution, we cannot detect line splitting in the $I$ spectrum, or
structure in the variation of $V/I$ with wavelength, from which to
obtain further information about field morphology.

%%%%%%%%%%%%%%%%%%%%%%%%%%%%%%%%%%%%%%%%%%%%%%%%%%%%%%%%%%%%%%%%%%%%%%%%%%%%%%%%
%%%%%%%%%%%%%%%%%%%%%%%%%%%%%%%%%%%%%%%%%%%%%%%%%%%%%%%%%%%%%%%%%%%%%%%%%%%%%%%%

\subsection{Other results (non-detections)}

%%%%%%%%%%%%%%%%%%%%%%%%%%%%%%%%%%%%%%%%%%%%%%%%%%%%%%%%%%%%%%%%%%%%%%%%%%%%%%%%%%%%%%%%%%%%%%%%%%%%%%%%%%%%%%%%%%%%%%%
\begin{table*}
\caption{White dwarfs in which weak fields may be present}
\label{Tab:weak_fields}
\begin{center}
\begin{tabular}{lllrrllcl}
\hline\hline
\multicolumn{2}{c}{Star names}  &Spectral&\te& $\log g$ &Field detected? & REF. & Field detected? & strength range \\
\multicolumn{2}{c}{}         & type    &(K)&          &(prev. work)   &      & (this work)    & (kG)           \\
\hline
WD\,0413$-$077  & 40 Eri B   & DA3   &17100& 7.95    & P &FVB03        & NA &  NA               \\
WD\,0446$-$789  & BPM 3523   & DA3   &23627& 7.69    & Y &AJN04        & Y  &  $-2.5$ to $-5.7$ \\
WD\,1105$-$048  & LTT 4099   & DA3   &15142& 7.85    & Y &AJN04, VBF06 & P  &  $-7.9$ to $3.3$  \\
WD\,1620$-$391  & CD--38 10980 & DA2 &24231& 8.07    & P &JAN07        & N  &  --               \\
WD\,2007$-$303  & LTT 7987   & DA4   &14454& 7.86    & P &JAN07        & N  &  --               \\
WD\,2039$-$202  & LTT 8189   & DA2.5 &19188& 7.93    & P &JAN07        & N  &  --               \\
WD\,2105$-$820  & LTT 8381   & DA6   &10794& 8.19    & Y &t.w.         & Y  &  $8.1$ to $11.4$  \\
WD\,2359$-$434  & LTT 9857   & DA5   &8544 & 8.44    & Y &AJN04        & Y  &  $3.1$ to $4.1$   \\
\hline
\end{tabular}
\tablebib{
Effective temperatures from \citet{Lajoieetal07} or \citet{Koesteretal09}. References for the magnetic field
detections are as follow: 
FVB03: \citet{Fabrikaetal03};
AJN04: \citet{Aznaretal04}; 
VBF06: \citet{Valyavinetal06}; 
JAN07: \citet{Jordanetal07}; 
t.w.: this work. 
}
\end{center}
\end{table*}
%%%%%%%%%%%%%%%%%%%%%%%%%%%%%%%%%%%%%%%%%%%%%%%%%%%%%%%%%%%%%%%%%%%%%%%%%%%%%%%%%%%%%%%%%%%%%%%%%%%%%%%%%%%%%%%%%%%%%%%

Two of the three observations of \object{WD\,2151$-$015} are different from zero
at a little more than the $2\,\sigma$ level, hence they do not represent
a significant detection. However, the presence in this star of a $2 -
4$~kG field cannot be ruled out. All the field measurements for all
the remaining stars lie within $2\sigma$ of zero field.

\section{A revised list of detections of weak magnetic fields in DA white dwarfs}

To set the results of our survey into a broader context, we have
compiled a list that includes all DA WDs in which, according to this
and previous work, a measurement of a non-zero longitudinal magnetic
field was obtained with an error bar $\sigma_{\bz} \la 2$\,kG.

\subsection{FORS1 archive measurements of longitudinal field}
The largest database of spectropolarimetric data that have reached a
sufficiently high S/N to detect weak fields is that included in the
FORS1 data archive. Most WD spectropolarimetric observations were
obtained in the context of dedicated surveys \citep[][and this
work]{Aznaretal04,Jordanetal07}. In addition to them, the FORS1 data
archive includes also four additional spectropolarimetric observations
of DA WDs that were obtained mainly for calibration purposes.

To produce a homogeneous dataset incorporating our current
understanding of how best to treat FORS1 spectropolarimetry, and to
examine the data to see if new reductions reveal any significant
fields missed in the earlier reductions, all FORS1 measurements have
been re-reduced following the same procedure adopted for the results
discussed in Sect.~\ref{Sect_New_Observations} \citep{Bagnuloetal12}.
As discussed at length in that article, these re-reductions are
expected to provide significantly improved field strengths and
(especially) uncertainties compared to the initial published
reductions. Our 70 ``new'' measurements from ``old'' FORS1 archive
data are reported in Table~\ref{Tab:rev_fld_meas} (published online),
which is organised in a similar way as Table~\ref{Tab:survey_obs},
with the omission of the $V$ magnitude and the insertion of a new
column which refers to the ESO programme ID of the observing run
corresponding the the observation. Values of \te\ and $\log g$ are
taken from \citet{Lajoieetal07,Koesteretal09,Giammicheleetal12}. Note
that all our new field determinations from data obtained by
\citet{Aznaretal04} have the opposite sign compared to their original
publication, to conform to the usual sign convention for the mean
longitudinal component of a stellar magnetic field (i.e, positive when
pointing to the observer).

The result of our re-evaluation of magnetic field measurements in WDs
from previously published FORS1 data is to confirm detections in three
stars (\object{WD\,0446$-$789}, \object{WD\,1105$-$048}, and
WD\,2359$-$434).  Note that these detections are based on rather
limited datasets. Each of these stars was observed only twice; in
WD\,0446$-$789 and WD\,2359$-$434, a field was detected in both
epochs, while in WD\,1105$-$048 the magnetic field was detected only
in one of the two observing epochs. All field detections are only at
the significance level of 3 to 6\,$\sigma$.

On the basis of \bz\ measurements significant at the 2\,$\sigma$ to
3\, $\sigma$ level, possible fields detections were reported for the
stars \object{WD\,1620--391}, \object{WD\,2007--303}, and
\object{WD\,2039--202}. In the new reductions, only one of these
measurements remains significant at slightly more than the $2 \sigma$
level. We consider that there is at present no firm evidence that any
of these stars posess detected kG fields.

\subsection{Other spectropolarimetric observations of weak-field WDs}

A literature search for additional WDs with a $\la 20$\,kG detected
longitudinal magnetic field returned only two stars:
\object{WD\,0413$-$077} = 40\,Eri\,B \citep{Fabrikaetal03}, and one
confirming observation for the field of WD\,1105$-$048
\citep{Valyavinetal06}.

The detection of WD\,0413$-$077 by \citet{Fabrikaetal03} is based on
an accumulation of measurements, most of which individually are only
barely significant.  Furthermore, those observations were carried out
with rather old spectropolarimeters designed in the late 1970s and
early 80s. For these reasons, the detection of the field in
WD\,0413$-$011 still requires confirming observations.

The confirming field detection in WD\,1105$-$048
\citep{Valyavinetal06} was obtained within the context of a survey of
five WDs (and 2 sdBs), which otherwise reported null results.

\subsection{A list of white dwarfs with weak magnetic fields}

The entries of Table~\ref{Tab:rev_fld_meas}, complemented with the WD
targets of the survey by \citet{Valyavinetal06} and the
observation of WD\,0413$-$077 by \citet{Fabrikaetal03}, constitute a
database that includes high-precision field measurements for 36 DA
WDs. From Table~3 we have excluded four WDs for which \bz\
measurements were obtained with error bars substantially larger than
2\,kG.  The mean error bar is $0.8$\,kG, which corresponds to a
typical (firm) field detection threshold of about 5\,kG. From this
database we have extracted a final list of suspected or confirmed
weak-field DA WDs that satisfy the condition $\vert\bz\vert <
20$\,kG. This list of stars (sorted by RA) is given in
Table~\ref{Tab:weak_fields}, which is organised as follows. Cols.~1 to
4 give star ID, spectral type, effective temperature and
gravity. Col.~5 contains a comment as to whether the field measurement
that appeared in the original papers would correspond to firm
detection (Y), or to a possible detection (P). The publications that
include the original field measurements are listed in col.~6. Col.~7
shows a note reporting whether according to our re-reduction of FORS1
data of Table~\ref{Tab:rev_fld_meas}, a field was firmly (Y), possibly
(P) or not (N) detected. The acronym NA (not applicable) used for
40~Eri\,B means that the original data were not obtained with FORS and
were not re-analysed. Col.~8 provides a new estimate of the range of
field variation as determined from the measurements available to date,
including the revisions reported in Table~\ref{Tab:rev_fld_meas}.

Table~\ref{Tab:weak_fields} shows that the newly-detected weak-field
white dwarf, WD\,2105$-$082, has been observed sufficiently often, and
with sufficiently high precision to fully confirm the existence of the
detected field, and to conclude that \bz\ is probably nearly constant
with time. Field detection in WD\,2359$-$434 appears reasonably
secure, and the two measurements obtained so far are consistent with a
field nearly constant with time. The hotter stars WD\,0446$-$789,
WD\,1105$-$048, and 40\,Eri\,B, appear to have variable fields.

\subsection{Stokes $I$ profiles from UVES archive
data}\label{Sect_UVES}

%%%%%%%%%%%%%%%%%%%%%%%%%%%%%%%%%%%%%%%%%%%%%%%%%%%%%%%%%%%%%%%%%%%%%%%%%%%%%%%%
\begin{figure} \rotatebox{270}{\scalebox{0.38}{
\includegraphics*[3cm,1cm][20cm,25cm]{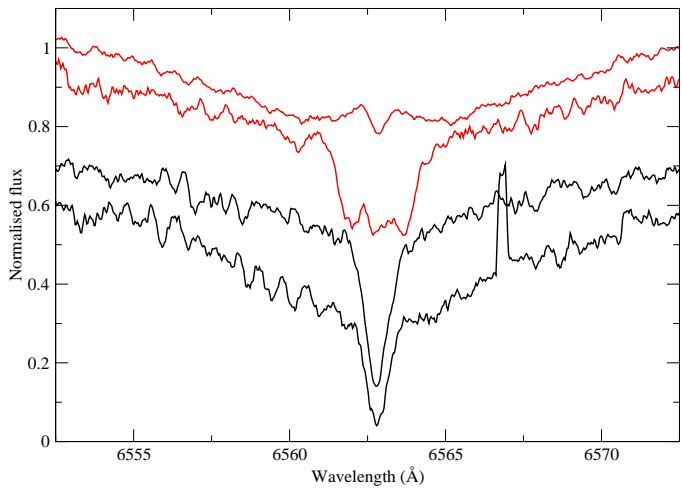}}}\\
\caption{\label{Fig:UVES_cool} H$\alpha$ cores of four cool WDs. Top
to bottom: WD\,2359$-$434 (magnetic, red line), WD\,2105$-$820
(magnetic, red line), WD\,1952$-$206, WD\,1826$-$045 (both
non-magnetic, according to our new FORS data). Spectra are normalised
to 1.0 at the edges of the window, then shifted vertically for display
purposes.}
\end{figure}

%%%%%%%%%%%%%%%%%%%%%%%%%%%%%%%%%%%%%%%%%%%%%%%%%%%%%%%%%%%%%%%%%%%%%%%%%%%%%%%%
\begin{figure} \rotatebox{270}{\scalebox{0.38}{
\includegraphics*[3cm,1cm][20cm,25cm]{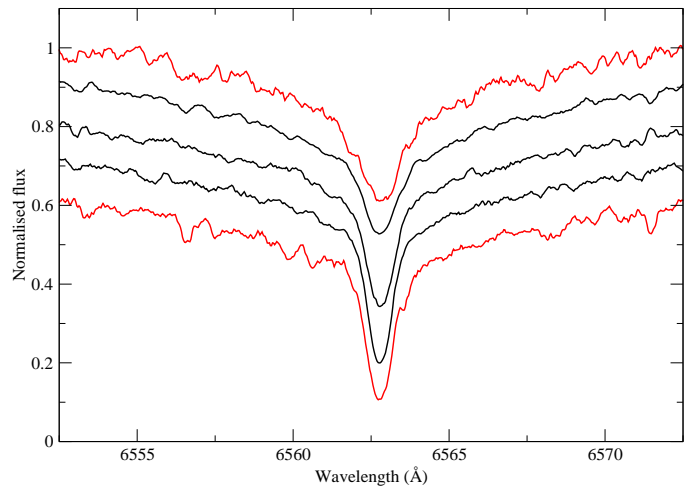}}}\\
\caption{\label{Fig:UVES_hot} H$\alpha$ cores of four hot white
dwarfs. Top to bottom: WD\,0446$-$789 (magnetic according to
Table~\ref{Tab:weak_fields}, red line), WD\,1620$-$391,
WD\,2039$-$202, WD\,2007$-$303 (all with longitudinal field consistent
with zero), WD\,1105$-$048 (possibly magnetic, red line).}
\end{figure}
%%%%%%%%%%%%%%%%%%%%%%%%%%%%%%%%%%%%%%%%%%%%%%%%%%%%%%%%%%%%%%%%%%%%%%%%%%%%%%%%

%%%%%%%%%%%%%%%%%%%%%%%%%%%%%%%%%%%%%%%%%%%%%%%%%%%%%%%%%%%%%%%%%%%%%%%%%%%%%%%%

Within the context of the SPY (Supernova Ia Progenitor surveY) project
\citep{Koesteretal09}, high-resolution UVES spectra were obtained at
two different epochs for each of the stars of
Table~\ref{Tab:weak_fields}, except for WD\,0413$-$077 = 40~Eri~B.  We
downloaded all these data from the UVES archive, to look for evidence
of Zeeman splitting in the H$\alpha$ line cores. The S/N in the
continuum around H$\alpha$ ranges from from 10 to over 100 per pixel
($=0.03$\,\AA). All spectra were smoothed with a running average over
nine pixels, i.e., the profiles were smoothed to an effective
resolution element of about 0.27~\AA, which is still small compared to
the FWHM of $\simeq 1$\,\AA\ for even the sharpest line cores.

Figure~\ref{Fig:UVES_cool} shows the H$\alpha$ line cores and inner
wings of the cooler stars of Table~\ref{Tab:weak_fields}, i.e.,
WD\,2359$-$434, and WD\,2105$-$820 (first and second spectra from the
top, respectively, in red). Both of these spectra show evidence of
Zeeman splitting. For comparison, Fig.~\ref{Fig:UVES_cool} also shows
the spectra of two non-magnetic cool white dwarfs of
Table~\ref{Tab:survey_obs}, \object{WD\,1826$-$045} and
\object{WD\,1952$-$206} (third and fourth lines from top,
respectively). The H$\alpha$ cores of the two latter WDs show no
significant excess width beyond that due to pressure broadening
\citep[see discussion by][]{Koesteretal98}.

As discussed in Sect.~\ref{Sect_Detection}, Zeeman splitting of the
H$\alpha$ core of WD\,2105$-$820 was already suspected by
\citet{Koesteretal98}; in the UVES data, with the smoothing adopted in
Fig.~\ref{Fig:UVES_cool}, the line core appears fairly clearly split
into the $\pi$ and two $\sigma$ components due to the Zeeman
effect. The magnetic field of this star is further discussed in
Sect.~\ref{Sect_Magnetic_Model}.

WD\,2359$-$434 has been discussed by \citet{Koesteretal98} and by
\citet{Koesteretal09}, who interpret the profile as showing a sharp
central $\pi$ component and two broad $\sigma$ components, which
suggests a rather non-uniform field with a mean value of $\bs \approx
100$~kG. A comparison between the the two available UVES spectra
(which were obtained four days apart) shows evidence of slight
variability. The small value of the ratio $\vert\bz\vert/\bs \sim
0.04$ suggests either that, if the field is roughly dipolar, we are
looking at it from nearly in the plane of the magnetic equator, or
that the field may be substantially more complex than a dipolar field,
perhaps somewhat like \object{WD\,1953$-$011} \citep{Valyavinetal08}.

Figure~\ref{Fig:UVES_hot} shows the H$\alpha$ cores for five hotter
stars of Table~\ref{Tab:weak_fields}.  The top spectrum is that of
WD\,0446$-$789, which according to Table~\ref{Tab:weak_fields} has a
field with \bz\ up to $-6$~kG. The H$\alpha$ line core appears to show
significant excess broadening compared to the others, probably due to
Zeeman splitting corresponding to a 20 to 30~kG field.  The two
available UVES spectra, separated by a time interval of about four
days, show only marginal signs of variability. The relatively high
ratio $\vert\bz\vert / \bs$ suggests that the star might have a
roughly dipolar morphology, with a polar field strength of order 30 to
40~kG.

The remaining four H$\alpha$ line cores of Figure~\ref{Fig:UVES_hot}
show no evidence of Zeeman splitting, and none of them show any
variation with time. Three of these spectra are those of stars which
have longitudinal magnetic field consistent with zero
(\object{WD\,1620$-$391}, WD\,2039$-$202, and
\object{WD\,2007$-$303}). The fourth unresolved core (the
lowest in the Figure) is that of WD\,1105$-$048, which according to
Table~\ref{Tab:weak_fields} has a field for which \bz\ ranges between
$-8$ and $+3$\,kG. With \bz\ this large, we would expect a \bs\ value
of the order of at least about 20\,kG, or even more. It is therefore
rather surprising that UVES spectra do not show any sign of Zeeman
broadening or splitting.

\section{A simple magnetic model for WD\,2105$-$820}\label{Sect_Magnetic_Model}

%%%%%%%%%%%%%%%%%%%%%%%%%%%%%%%%%%%%%%%%%%%%%%%%%%%%%%%%%%%%%%%%%%%%%%%%%%%%%%%%
%\begin{figure}
%\scalebox{0.44}{
%\includegraphics*[0.9cm,5.3cm][21.5cm,25cm]{Figure_UVES.ps}}\\
%\caption{\label{Fig:UVES} WD\,2105$-$820.
%The core of H$\alpha$ split by Zeeman effect. The original UVES spectrum was
%smoothed with a boxcar smoothing of 9 pixels (pixel bin is 0.03\AA)}
%\end{figure} 
%%%%%%%%%%%%%%%%%%%%%%%%%%%%%%%%%%%%%%%%%%%%%%%%%%%%%%%%%%%%%%%%%%%%%%%%%%%%%%%%

WD\,2105$-$820 is the only kG MWD for which there exists a sufficiently
large number of magnetic field measurements to allow us to start simple
modelling.

Four of our longitudinal field measurements were obtained during one
week, and the fifth one about one month later. During this time
interval, the field shows at most marginal evidence of variability,
and the observed fluxes show none.

Mean field modulus measurements made by \citet{Koesteretal98} from
CASPEC observations of excess H$\alpha$ line core broadening (which we
can now safely ascribe to the Zeeman effect) yield a field strength
\bs\ of $43 \pm 10$~kG. Although all three observations have very low
S/N, it appears that the three H$\alpha$ profiles show similar Zeeman
broadening. The first two measurements were obtained on 1995 July 13,
and the last on 1996 July 29, i.e., about a year later (D. Koester,
private communication). In addition, two further high-resolution
spectra of this star containing H$\alpha$ were obtained with UVES for
the SPY project \citep{Koesteretal09}, one on 2002 May 29, and one on
2003 May 13 (see Sect.~\ref{Sect_UVES}). \citet{Koesteretal09} remark
that the excess broadening of H$\alpha$ in these spectra is very
similar in width to that observed in the three older spectra, and thus
they find no evidence that \bs\ has changed. 
The S/N of the 2003 spectrum is too low to
make possible an accurate determination of \bs, but in the 2002
spectrum, the H$\alpha$ core appears to show the $\pi$ and two $\sigma$
components clearly, with a $\sigma - \sigma$ separation of $\sim
1.7$\,\AA, corresponding to $\bs = 42 \pm 3$~kG. Since the available
Stokes~$I$ measurements were obtained over a span of eight years, we
conclude that the observed mean field modulus of WD\,2105$-$820 does
not change much even over a time scale of a decade.

Assuming that the star's magnetic model can be described in terms of
the oblique rotator model, which seems to be generally true of MWDs
that have been modelled in detail \citep{Landstreet92,Kulebietal09b},
these results indicate that either (1) the stellar rotation period is
much longer than one year (or possibly shorter than the integration
time of the observations), or that the magnetic structure is such that
the observed field does not vary much as the star rotates, i.e., (2)
the field is roughly symmetric about the rotation axis, or (3) the
rotation axis is nearly aligned to the line of sight.

We note that none of the variable MWDs with known periods discussed by
\citet{Schmidtetal91} or \citet{Kawkaetal07} \citep[see also Table~2
of][]{Landstreet92} have rotation periods longer than 18 days (and
only one has a rotation period significantly shorter than 1~hr)).
Thus we consider the hypothesis of a field
approximately symmetric about the star's rotation axis, or possibly of a
stellar rotation axis nearly parallel to the line of sight. 

We furthermore note that the value of the ratio $\bz / \bs \approx
0.22$ is a strong indicator of a rather simple magnetic field
structure (a much smaller value is expected for complex fields such as
those of solar-type stars). In particular, the value of this ratio is
consistent with a dipolar morphology \citep{Landstreet88,
  Schmidtetal91}, which we adopt as a magnetic model for
WD\,2105$-$820. We note that in their modelling of DAH stars with
stronger fields, \citet{Kulebietal09b} frequently obtained better fits
to their (time-averaged) $I$ spectra with decentred dipoles than with
centred ones, but for the weak field of WD2105$-$820 we do not have a
strong constraint on possible decentring in the available data.

If we assume that the magnetic field is symmetric about the star's
rotation axis, then the dipolar axis must be parallel to the stellar
rotation axis. Using Eqs.~(1), (2), (6), (8), and (21) of
\citet{Hensbergeetal77} (setting the limb darkening coefficient to 1,
$\bz={\rm const}=10$\,kG, $\bs={\rm const}=43$\,kG), we find that the
observations are consistent with a simple centred dipole with a polar
field strength of $\sim 56$\,kG, and magnetic axis parallel to the
rotation axis inclined at about $\sim 68\degr$ with respect to the
line of sight. If we assume a rotation axis parallel to the line of
sight, then magnetic field observations are explained again by a
dipole with field strength at the pole of $\sim 56$\,kG, but with
dipole axis tilted at $\sim 68\degr$ with respect to the rotation axis
(which is parallel to the line of sight). Note that the field models
obtained in the two cases are the same; the only difference between
the models is that the inclination of the rotation axis to the line of
sight $i$, and the obliquity angle between the rotation and dipole axes
$\beta$, have been exchanged.

\section{Discussion and conclusions}
The database that we have considered includes 20 hot DA stars
(generally spectral type DA1 to DA4, $\te \ga 14000$\,K) and 15 cool
DA stars (spectral type DA5 to DA8; $\te \la 14000$\,K). (We
  omit 40 Eri B from our sample, as we have no data to confirm
the field detected, and the star was not observed in a survey of known
size.) Since there are two firmly detected MWDs in each of the hot and
cool samples, we conclude that detection rates are about 10\,\% for
the hot sample, and 13\,\% for the cool sample. The small size of the
sample and the small number of detections set a serious limit to
accuracy of these frequency estimates. Using the Wilson 95\,\%
confidence limits \citep{Wilson27}, the field detection rate in hot
WDs could be anywhere between 2.8 and 30\,\%, while the field
detection rate in cool DA WDs lies between 3.7 and 38\,\%. In
conclusion, the data currently available are consistent with the
hypothesis that weak magnetic fields occur with the same frequency in
hot and cool DA WDs.  Globally, the detection of four weak magnetic
fields from a total sample of 36 WDs makes it quite clear that the
probability of finding a weak field in a DA WD is neither negligible,
nor close to 1; at the 95\% confidence limits, the probability lies
between 4 and 25\%.  Therefore, it appears that the probability of
detecting a $\sim 10$\,kG field in a WD is comparable to the
probability of detecting a magnetic field with strength in the range
100\,kG -- 500\,MG, which is $\sim 10$\,\%.

Re-addressing some of the questions posed in Sect.~1, it appears now
that $\sim 10$\,kG longitudinal fields are not ubiquitous in WDs
lacking stronger fields, nor do fields seem to die away at this level.
Furthermore, we have not found any significant
difference between field detection rates in cool, old WDs and field
detection rate in hot, young WDs. Studying these questions further
will require substantially larger samples of precise field
measurements than those available now.

The results of this paper highlight the need for (1) further field
measurement of the MWDs already detected in this low-field regime, to
fully confirm the reported detections, and to provide data on possible
variabilty in order to characterise the field structures observed; (2)
an extended high-precision survey of magnetic fields in hot and cool
WDs, aimed at refining the frequency of occurrence of weak fields in
the range studied here; and (3) a still deeper survey, using long
integrations, to reach even weaker fields (note that standard errors
of 300 -- 500\,G are already achieved in a number of stars with
integrations of mostly less than 30\,min). It will also be interesting
to discover whether the morphologies of the fields of kG MWDs are
often roughly symmetric about the rotation axis, as seems to be the
case for WD\,2105$-$820 and as frequently happens for MWDs with
stronger fields. All of these goals are within reach of observing
programmes on the VLT with FORS2, although they would be very
difficult on smaller telescopes.

After this paper was accepted, S. Vennes communicated to us the
results of a survey of magnetic fields in a sample of 58 high proper
motion white dwarfs \citep{KawkaVennes12}. The stars of their survey
are complementary to the two samples discussed in our paper. Our hot
sample contains stars with typical cooling ages of 300 Myr or less,
and our cool sample WDs typically have cooling ages of 300 -- 1000
Myr, while the sample of Kawka \& Vennes is made up largely of stars
with cooling ages above 1 Gyr. Because the WDs observed by Kawka \&
Vennes are both cooler and typically 2--3~mag fainter than those of
our samples, their median standard error of field measurement is about
3~kG, compared to about 800~G for our sample. They are thus sensitive
mainly to \bz\ fields larger than 10--20 kG, just above the \bz\ range
of greatest interest to our study. However, their results seem to be
significantly different from ours, as they find a probability of field
detection of the order of 1 -- 2\% per decade of field strength, while
the samples discussed by us suggest probabilities of the order of 10\%
per decade in the weak-field limit. Further observations will be
needed to determine if this difference is real. If the difference is
indeed real, it may be an evolutionary effect of field decay with
time, or a real increase in probability as we probe smaller and
smaller field strengths.

\begin{acknowledgements}
  We thank the referee, Prof. Gary Schmidt, for a careful reading of
  the manuscript and for helpful comments.  JDL and GAW acknowledge
  financial support from the Natural Sciences and Engineering Research
  Council of Canada.

\end{acknowledgements}

\Online
\begin{longtable}{lllrrrrrr@{$\,\pm\,$}lc}
\caption{\label{Tab:rev_fld_meas}
Revised \bz\ field strength values for all magnetic field measurements of potential kG
field DA white dwarfs, obtained from H Balmer lines only}.
\\
\hline\hline
%---------------------
\multicolumn{2}{l}{Star names} &  %1,2
Spec.                          &  %3
\te                            &  %4
$\log g$                       &  %5
ESO Pr. ID                     &  %6 
\multicolumn{1}{c}{MJD}        &  %
$t_{\rm int}$                     &  %8
\multicolumn{2}{c}{\bz}         &  %9,10
field                          \\
&&type& (K) &&&&
(s)                            &  %8
\multicolumn{2}{c}{(G)}         & %9,10
detected?                      \\ %11                         
\hline
\endfirsthead
\caption{Table 3., continued}\\
\hline\hline
\multicolumn{2}{l}{Star names} &  %1,2
Spec.                       &  %3
\te                            &  %4
$\log g$                       &  %5
ESO Pr. ID.                    &  %6 
\multicolumn{1}{c}{MJD}        &  %7
$t_{\rm int}$                     &  %8
\multicolumn{2}{c}{\bz}        &  %9,10
field                         \\  %11
&&type& (K) &&&&
(s)                            &  %8
\multicolumn{2}{c}{(G)}        & %9,10
detected?                     \\ %11
\hline
\endhead
\hline
\endfoot
   WD\,0135-052 & NLTT 5460       &  DA7  &  7273 & 7.85 & 070.D-0259  &  52608.097  & 1828 &$  -555 $& 530  &N\\ 
   WD\,0227+050 & GJ 100.1        &  DA3  & 18887 & 7.84 & 070.D-0259  &  52637.120  & 2292 &$   671 $& 590  &N\\ 
                &                 &       &       &      &             &  52669.062  & 2292 &$  -524 $& 580  & \\ 
   WD\,0310-688 & GJ 127.1        &  DA3  & 15658 & 8.09 & 070.D-0259  &  52695.054  & 1680 &$    85 $& 420  &N\\ 
   WD\,0346-011 & GD 50           &  DA1  & 41196 & 9.15 & 070.D-0259  &  52637.176  & 2800 &$  1307 $& 3540 &N\\ 
                &                 &       &       &      &             &  52674.078  & 2800 &$ -1818 $& 3780 & \\ 
   WD\,0446-789 & BPM 3523        &  DA3  & 23627 & 7.69 & 070.D-0259  &  52609.229  & 2800 &$ -2548 $& 820  &Y\\ 
                &                 &       &       &      &             &  52668.087  & 2800 &$ -5670 $& 935  & \\ 
   WD\,0612+177 & NLTT 16280      &  DA2  & 25312 & 7.94 & 070.D-0259  &  52609.274  & 2800 &$ -1445 $& 725  &N\\ 
                &                 &       &       &      &             &  52672.079  & 2800 &$  -341 $& 730  & \\ 
   WD\,0631+107 & KPD 0631+1043   &  DA2  & 26718 & 7.87 & 070.D-0259  &  52700.130  & 2800 &$ -1182 $& 1100 &N\\ 
                &                 &       &       &      &             &  52702.125  & 2800 &$   335 $& 1095 & \\ 
   WD\,0839-327 & LTT 3218        &  DA6  &  9318 & 7.99 & 070.D-0259  &  52608.319  & 1870 &$   315 $& 250  &N\\ 
   WD\,0859-039 & WD\, J0902-041  &  DA2  & 23731 & 7.79 & 070.D-0259  &  52674.227  & 2320 &$  -148 $& 770  &N\\ 
                &                 &       &       &      &             &  52696.219  & 2320 &$ -1168 $& 735  & \\ 
   WD\,1042-690 & NLTT 25239      &  DA3  & 21012 & 7.93 & 070.D-0259  &  52668.351  & 2562 &$ -1008 $& 755  &N\\ 
                &                 &       &       &      &             &  52674.273  & 2562 &$   503 $& 755  & \\ 
                &                 &       &       &      &             &  52695.301  & 2562 &$    27 $& 575  & \\ 
   WD\,1105-048 & NLTT 26379      &  DA3  & 15142 & 7.85 & 070.D-0259  &  52641.351  & 1948 &$    59 $& 580  &P\\ 
                &                 &       &       &      &             &  52669.305  & 1948 &$  3341 $& 655  & \\
   WD\,1202-232 & EC12028-2316    &   DA  &  8615 & 8.04 & 073.D-0356  &  53144.146  & 2000 &$  -392 $& 605  &N\\ 
                &                 &       &       &      &             &  53147.179  & 2000 &$  -383 $& 435  & \\ 
                &                 &       &       &      &             &  53150.997  & 2000 &$  -343 $& 440  & \\ 
   WD\,1327-083 & G 14-58         &  DA4  & 13823 & 7.80 & 073.D-0356  &  53151.033  & 1740 &$  -201 $& 480  &N\\ 
                &                 &       &       &      &             &  53153.068  & 1710 &$   149 $& 485  & \\ 
   WD\,1334-678 & LTT 5267        &  DA6  &  8769 & 7.93 & 073.D-0516  &  53134.050  & 1384 &$  4017 $& 3285 &N\\ 
                &                 &       &       &      &             &  53137.010  & 1384 &$ -5021 $& 4210 & \\ 
   WD\,1425-811 & LTT 5712        &  DA6  & 12098 & 8.21 & 073.D-0516  &  53137.044  &  960 &$   488 $& 1360 &N\\ 
   WD\,1620-391 & CD-38 10980     &  DA2  & 24231 & 8.07 & 069.D-0210  &  52383.426  &  240 &$   223 $& 775  &N\\ 
                &                 &       &       &      &             &  52383.431  &  300 &$   703 $& 1110 & \\ 
                &                 &       &       &      & 073.D-0356  &  53136.301  & 1022 &$   188 $& 335  & \\ 
                &                 &       &       &      &             &  53143.322  & 1022 &$   184 $& 500  & \\ 
                &                 &       &       &      &             &  53147.255  & 1022 &$   -12 $& 315  & \\ 
                &                 &       &       &      &             &  53151.070  & 1022 &$   -48 $& 420  & \\ 
   WD\,1733-544 & LTT 6999        &  DA8  &  6165 & 7.23 & 073.D-0516  &  53199.178  & 1664 &$  4104 $& 4390 &N\\ 
   WD\,1826-045 & LTT 7347        &  DA6  &  9057 & 7.91 & 073.D-0516  &  53193.179  & 1920 &$ -2705 $& 1535 &N\\ 
   WD\,1845+019 & LAN 18          &  DA2  & 29384 & 7.81 & 073.D-0356  &  53131.395  & 2100 &$    76 $& 855  &N\\ 
                &                 &       &       &      &             &  53136.389  & 2100 &$    99 $& 755  & \\ 
   WD\,1919+145 & GD 219          &  DA5  & 14430 & 8.06 & 073.D-0356  &  53132.324  & 2100 &$ -1451 $& 790  &N\\ 
                &                 &       &       &      &             &  53136.351  & 2100 &$  -953 $& 760  & \\ 
   WD\,1952-206 & LTT 7873        &  DA6  & 13184 & 7.82 & 073.D-0516  &  53251.088  & 2840 &$   530 $& 1180 &N\\ 
   WD\,2007-303 & LTT 7987        &  DA4  & 14454 & 7.86 & 067.D-0306  &  52076.437  &  200 &$  2058 $& 2670 &N\\ 
                &                 &       &       &      & 073.D-0356  &  53132.382  & 3600 &$   501 $& 360  & \\ 
                &                 &       &       &      &             &  53138.373  & 1800 &$  -490 $& 400  & \\ 
   WD\,2014-575 & RE J2018-572    &  DA2  & 27465 & 7.94 & 073.D-0356  &  53140.360  & 2100 &$   730 $& 1075 &N\\ 
                &                 &       &       &      &             &  53184.273  &  700 &$ -5213 $& 2235 & \\ 
                &                 &       &       &      &             &  53185.107  & 2100 &$  -697 $& 1230 & \\ 
   WD\,2039-202 & LTT8189         &  DA3  & 19188 & 7.93 & 060.A-9203  &  53869.443  &  851 &$ -4367 $& 1915 &N\\ 
                &                 &       &       &      & 073.D-0322  &  53148.420  &  532 &$   944 $& 770  & \\ 
                &                 &       &       &      & 073.D-0356  &  53143.362  & 1800 &$  -288 $& 645  & \\ 
                &                 &       &       &      &             &  53167.393  & 1800 &$   604 $& 400  & \\ 
   WD\,2105-820 & LTT 8381        &  DA6  & 10794 & 8.19 & 073.D-0516  &  53192.269  & 1760 &$  9274 $& 1375 &Y\\ 
                &                 &       &       &      &             &  53193.278  & 1760 &$ 11423 $& 995  & \\ 
                &                 &       &       &      &             &  53197.294  &  880 &$  8173 $& 1630 & \\ 
                &                 &       &       &      &             &  53199.317  &  880 &$  9130 $& 1490 & \\ 
                &                 &       &       &      &             &  53227.209  & 1760 &$  9770 $& 845  & \\ 
   WD\,2115-560 & LTT 8452        &  DA6  &  9625 & 8.01 & 073.D-0516  &  53199.342  & 1664 &$ -1367 $& 1065 &N\\ 
                &                 &       &       &      &             &  53227.238  & 1664 &$   132 $& 955  & \\ 
   WD\,2149+021 & G 93-48         &  DA3  & 17360 & 7.93 & 073.D-0356  &  53183.278  & 2088 &$  -875 $& 675  &N\\ 
                &                 &       &       &      &             &  53196.346  & 2088 &$   354 $& 575  & \\ 
                &                 &       &       &      &             &  53222.200  & 2088 &$     2 $& 540  & \\ 
   WD\,2151-015 & NLTT 52306      &  DA6  &  9194 & 7.97 & 073.D-0516  &  53240.174  & 1840 &$  3941 $& 1910 &N\\ 
                &                 &       &       &      &             &  53251.124  & 1840 &$ -2027 $& 950  & \\ 
                &                 &       &       &      &             &  53252.120  & 1840 &$  -688 $& 1660 & \\ 
   WD\,2211-495 & RE J2214-491    &   DA  & 62236 & 7.54 & 073.D-0356  &  53140.401  & 1610 &$   819 $& 1095 &N\\ 
                &                 &       &       &      &             &  53185.246  & 1610 &$  -576 $& 1205 & \\ 
   WD\,2333-049 & G 157-82        &  DA6  & 10608 & 8.04 & 073.D-0516  &  53274.201  & 1704 &$  5102 $& 5550 &N\\ 
   WD\,2359-434 & LTT 9857        &  DA5  &  8544 & 8.44 & 070.D-0259  &  52583.025  & 2188 &$  4097 $& 840  &Y\\ 
                &                 &       &       &      &             &  52608.056  & 2188 &$  3090 $& 510  & \\ 
\end{longtable}
\listofobjects
\end{document}